\documentclass[copyright,creativecommons]{eptcs}
\providecommand{\event}{MeCBIC 2010} 

\usepackage[T1]{fontenc}
\usepackage{amsfonts}
\usepackage{graphicx}
\usepackage{epsfig}
\usepackage{wrapfig}
\usepackage{hyperref}
\usepackage{url}
\usepackage{amssymb}
\usepackage{amsmath}
\usepackage{calc}
\usepackage{subfig}
\usepackage{xcolor}

\newtheorem{definition}{Definition}[section]

\newcommand{\oc}{$\mbox{O}_c$}
\newcommand{\ob}{$\mbox{O}_b$}
\newcommand{\oy}{$\mbox{O}_y$}
\newcommand{\pc}{$\mbox{P}_c$}
\newcommand{\pb}{$\mbox{P}_b$}

\newcommand{\St}{{\mathbb S}}

\newcommand{\bbbn}{\mathbb{N}}
\newcommand{\bbbz}{\mathbb{Z}}

\newcommand{\me}{\textit{ME}}

\title{Multiscale Bone Remodelling with Spatial P Systems}
\author{Diletta Cacciagrano, Flavio Corradini, Emanuela Merelli, Luca Tesei
\institute{School of Science and Technology, Computer Science Division\\
University of Camerino, Camerino, Italy\thanks{This work was partially supported by the Italian
FIRB-MIUR LITBIO: \textit{Laboratory for Interdisciplinary Technologies in
Bioinformatics}.}\\}
\email{\{name.surname\}@unicam.it}
}
\def\titlerunning{Bone Remodelling}
\def\authorrunning{D. Cacciagrano, F. Corradini, E. Merelli \& L. Tesei}
\begin{document}
\maketitle

\begin{abstract}
Many biological phenomena are inherently multiscale, i.e.\ they are characterized by interactions
involving different spatial and temporal scales simultaneously. Though several approaches have been proposed to provide ``multilayer'' models, only Complex Automata, derived from Cellular Automata, naturally embed spatial information and realize multiscaling with well-established inter-scale integration schemas. Spatial P systems, a variant of P systems in which a more geometric concept of space has been added, have several characteristics in common with Cellular Automata. We propose such a formalism as a basis to rephrase the Complex Automata multiscaling approach and, in this perspective, provide a 2-scale Spatial P system describing bone remodelling. The proposed model not only results to be highly faithful and expressive in a multiscale scenario, but also highlights the need of a deep and formal expressiveness study involving Complex Automata, Spatial P systems and other promising multiscale approaches, such as our shape-based one already resulted to be highly faithful.
\end{abstract}

\section{Introduction}
Nowadays, it is possible to observe biological systems in great detail: with a light microscope one
can distinguish the compartments of a human cell, and with an electron microscope one can even see
very small details such as proteins. At the same time, models for describing and simulating biological systems have comparable resolution regimes and work on different spatial and temporal scales.

\begin{figure}
\begin{center}
\includegraphics[width=11cm]{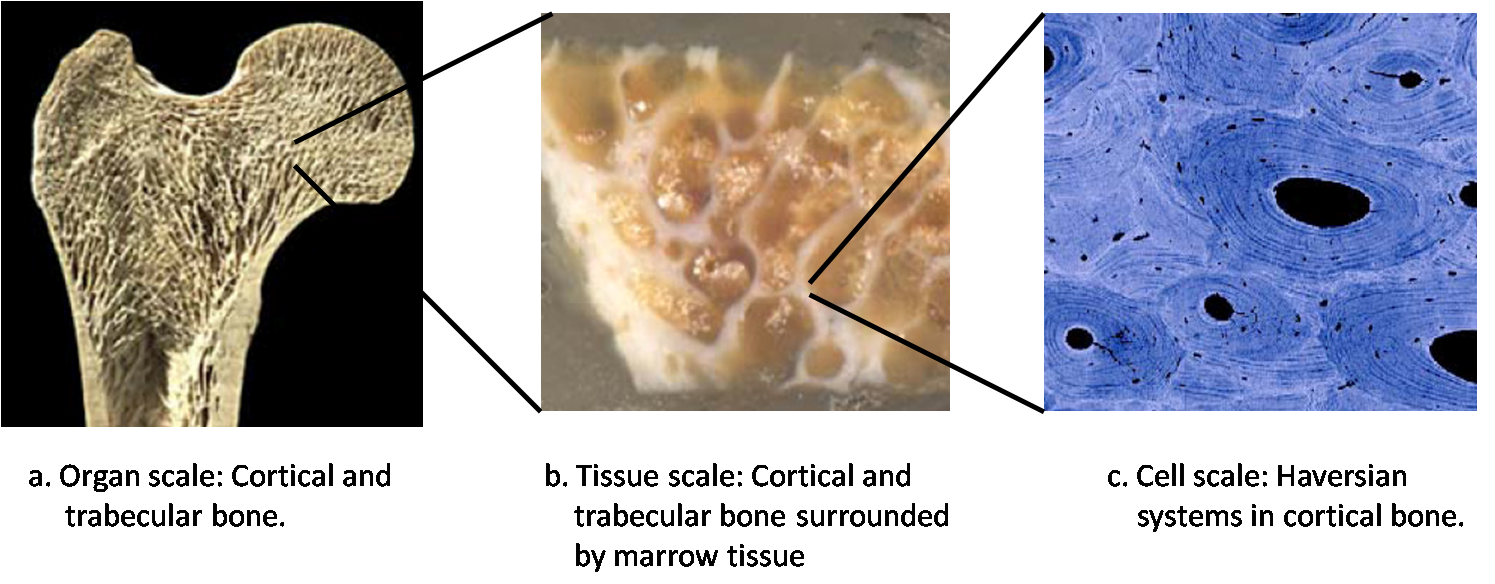}
\end{center}
\caption{Multiscale view of a human femur.}
\label{multi}
\end{figure}

Actually, a characteristic of biological complexity is the {\em intimate connection} that exists between such scales. The {\em bone remodelling} \cite{frost90}, concerning the continuous replacement of old bone by new tissue, is just an exemplar multiscale phenomenon, where macroscopic behaviour (at organ and tissue scale) and microstructure (at cell scale) strongly influence each other (see Fig.~\ref{multi}).

At tissutal scale, two macroscopically different bone tissue types are distinguished: the {\em cortical} one - which is a rather dense tissue although it is penetrated by blood vessels through a network of canaliculi - and the {\em trabecular} one - which is porous and primarily found near joint surfaces, at the end of long bones and within vertebrae. It is well-known that external mechanical loading allows bone to adapt its structure in response to the mechanical demands; in particular, collagen fibers in  bone tend to align with maximum stresses in many bones and greatly increase their load-carrying capacity without increasing mass, thus improving structural efficiency.

At cellular scale\footnote{For a more detailed description, see \url{http://courses.washington.edu/bonephys/physremod.html}.}, two main kinds of cells, namely {\em osteoclasts} ($\mbox{O}_c$) and {\em osteoblasts} ($\mbox{O}_b$), closely collaborate in the remodelling process in what is called a {\em Basic Multicellular Unit} (BMU). The remodelling process begins at a quiescent bone surface (either cortical or trabecular) with the appearance of $\mbox{O}_c$s, which attach to the bone tissue matrix, form a ruffled border, create an isolated microenvironment, acidify it and dissolve the organic and inorganic matrices of the bone.

Briefly after this resorptive process stops, $\mbox{O}_b$s appear at the same surface site, deposit osteoid and mineralize it. Some $\mbox{O}_b$s are encapsulated in the osteoid matrix and differentiate to {\em osteocytes} ($\mbox{O}_y$). Remaining $\mbox{O}_b$s continue to synthesize bone until they eventually stop and transform to quiescent {\em lining cells} ($\mbox{L}_c$) that completely cover the newly formed bone surface and connect with the $\mbox{O}_y$s in the bone matrix through a network of canaliculi.

Bone remodelling has been subject of extensive studies in many fields of research: much of the proposed approaches are based on reduction - i.e.\ isolating the various components to unravel their individual behaviour - without taking into account how mechanical forces are translated to structural adaptation of the internal cellular architecture \cite{BSL03,HR03}, while other approaches relate density changes in bone directly to local strain magnitudes, abstracting from the underlying cellular processes (i.e.\ morphology and metabolic activity) \cite{ch76,ch76bis,hui00}. On the contrary, the actual knowledge about bone remodelling shows several gaps at different resolution degrees {\em at the same time} \cite{vic2008}. For instance:
\begin{itemize}
\item[-] {\em (Tissue level)} There are some questions as to whether the orientation of collagen fibers in bone occurs through functional adaptation as the bone is being remodelled or is under genetic influence during development.
\item[-] {\em (Cell level)} BMU existence indicates that a coupling mechanism must exist between formation and resorption, i.e.\ among $\mbox{O}_b$s and $\mbox{O}_c$s. However, the nature of this  mechanism is not known.
\item[-] {\em (Cell-Tissue level)} It is not so clear how mechanical forces can be expressed in cell activity and whether they are enough to explain remodelling. The current concept is that the bone architecture is also controlled by local regulators and hormones (mainly insulin-like growth factors, cytokines interleukin-1, interleukin-6 and RANKL) and that both local mechanical and metabolic signals are detected from $\mbox{O}_y$s. Whether this is true remains to be proven.
\end{itemize}

\subsection{Motivations and contribution of the paper}
Computational science is becoming more ambitious by moving beyond the traditional approach of modelling individual isolated systems,
towards integrated  systems having numerous mutually interacting components. {\em Multiscale models} just fall in this category as {\em coupled models}, where coupling is often supported by domain specific (only in few cases, slightly more general) solutions. Although multiscale models allow a high expressive description of a system, they are not always more ``faithful''
\footnote{Informally, a model describing a complex system is considered ``faithful'' whenever the abstract representation it provides is so
close to the real system that allowing (the most of) system properties to be correctly inferred from it.} than  single-scale ones. In fact, such a ``faithfulness'' is invalidated whenever spatial information is fully ignored and  approximation techniques are used to integrate the different components.

{\bf The need of spatial information.} Space is fundamental to add ``faithfulness'' to biological models (not only multiscale) \cite{iccs2010,cs2bio2010,Pru2006,god2010,gia03,kur94}. In the case of bone remodelling,
for instance, the coalescence process, i.e.\ the formation of $\mbox{O}_c$s, is possible only when a {\em sufficient}
number of pre-osteoclasts ($\mbox{P}_c$) are available and {\em quite close} to each other.
The need of making explicit space in both single and multiscale models has been already stressed - just taking into account
bone remodelling -  in  \cite{cs2bio2010},  where a cell and tissue scale model of the phenomenon have been defined and
integrated in terms of shapes equipped with perception, interaction and movement capabilities. Space has been there exploited to
better understand the blurry synergy between mechanical and metabolic factors triggering bone remodelling, both in qualitative
and in quantitative terms.

Obviously, bone remodelling is not the only biological phenomenon where space plays a crucial role. For instance, cytoplasm (of even the simplest cell) and enzymes are other excellent examples. The first contains many distinct compartments. In each compartment, localization of molecules can be influenced in many different ways, such as by anchoring to structures like the plasma membrane or the cytoskeleton. The latter, acting in the same pathway, are often found co-localized; as the product of one reaction is the substrate for the next reaction along the pathway, this co-localization increases substrate availability and concomitantly enhances catalytic activity by giving rise to increased local concentration of substrates.

{\bf The need of uniform modelling components.} A multiscale model can be more or less ``faithful'' according to what ``single-scale'' components are taken into account (for each scale) and how they are ``homogenized'' (i.e.\ integrated) \cite{hoeuno,hoedue,iccs2010,cs2bio2010,acri2010}. Homogenization is a very delicate and complex task which can often lead to loss of information between scales - in particular when both ``single-scale'' components are specifically heterogeneous (so that needing approximation techniques to be integrated) and inter-scale synergies are particularly complex (so that admitting different homogenization schemas).

Just to give an idea of such a complexity, the simplest systems to homogenize are only those in which a fine scale model can be coarsened (averaged) to produce key data for a coarser level model. For instance, an atomistic model of a metal can be used to evaluate, ab initio, its shear and bulk moduli: having found these two parameters, the coarse grained model (the equations of elasticity) can then be solved, without further reference to the fine scale atomistic model.

On the contrary, in most of the cases the microscale problem depends on the macroscale variables, hence systems do not decouple so easily and different homogenization schemas can arbitrarily been applied. Consider, for instance, the fluid flow through a vessel network, where
the fine scale structure of the flow depends on the (coarse scale) pressure gradient. In this case, two different homogenization schemas (respectively {\em Matched asymptotic expansion method} and  {\em Multiple scale method}) could be  taken into account.
According to the first, a fine scale could correspond to the flow through a single vessel, an intermediate scale to vessels acting as
discrete network, and a coarse scale to vessels acting as an effective porous material. According to the latter, a fine scale
could correspond to the flow through a periodic network of vessels, and a coarse scale to vessels acting as an effective porous material.

As a consequence, a high uniformity degree among ``single-scale'' components implies the possibility of defining well-established homogenization rules and increasing the ``faithfulness'' of a multiscale model in the whole.

This aspect has been discussed in \cite{cs2bio2010} where the shape-based approach, adopted for defining both tissutal and cellular scales of bone remodelling, is crucial for easily achieving their integration. In detail, the coupling is realized discretizing the trabecular tissue as a grid of cells, associating a shape-based (and very detailed) BMU cellular model to each cell and dynamically alternating in it shapes, taken from a finite family of basic shapes associated to specific mineralisation density values, in according to the density value computed by the underlying cellular model.

{\bf Contribution of the paper.} {\em Complex Automata} (CxA) \cite{HoekstraFCC08} are very close to what we consider to be a ``faithful'' multiscale modelling paradigm (see Section~\ref{cxa}). In fact, they naturally embed spatial information and realize multiscaling on uniform components (namely {\em Cellular Automata} (CA) \cite{new66}) by different and well-established integration schemas. Any CxA building block is composed of a finite grid of cells, where each cell has an associated state taken from a finite set of different states. All the cells change state in accordance with a rule, which is characteristic of the particular block. The rule is deterministic and ``local'' - in the sense that the new state of a cell is determined only on the basis of the previous states of cell itself and of nearby cells. For this reason, CxA rely on a rigid concept of space, being directly inherited from the CA notation.

This aspect has been already taken into account in \cite{acri2010}, where a 2-component CxA, based on a micro-macro integration scheme, has been first defined for describing bone remodelling and then executed in {\sc BioShape}\footnote{{\sc BioShape} modelling and simulation environment can be found at \url{http://cosy.cs.unicam.it/bioshape}}. In particular, the possibility in {\sc BioShape} to associate to each shape  {\em its own} physical movement law (which can be different from that one associated to any neighbour) already raised the need of investigating the CxA expressiveness about spatial heterogeneity.
\par\smallskip
{\em Spatial P systems} \cite{sps} (SP), a variant of P systems where a more geometric concept of space has been added, have several characteristics in common with CA. SPs (see Section~\ref{sp}) embed a geometric grid-based 2D concept of space. Cells in a SP contain objects and can be organized in a hierarchy of membranes. Interactions in a cell occur between objects and are described with evolution rules defined in the membrane to which every cell belongs. The evolution rules are space-based, i.e.\ their applicability can depend on the presence of certain objects in the same cell or in neighbour cells. Moreover, each cell of an SP can accommodate any number of objects, but only one of the so-called Mutually Exclusive ({\me}) objects. This feature is useful to represent particular kind of objects that are larger than normal ones and that create a context in a cell such that other objects with the same characteristics cannot be created. {\me} objects, in conjunction with membranes, can also be used to represent physical obstacles and to enclose certain compartments in which specific activities occur. However, differently from CxA, SPs lack of well-established integration schemas.
\par\smallskip
Bone remodelling just becomes the pivot for facing both approaches. The SP paradigm could easily rephrase the CxA proposed in \cite{acri2010}, being both SPs and CxA able to describe spatial lattices and to model different spatial and temporal layers. For this reason, the SP paradigm is instead exercised stressing directly those modelling features which seem to characterize only SPs but not CxA (see Section~\ref{cs2}). In this perspective, the cellular scale is depicted by explicitly modelling local regulators, {\ob}s, {\oc}s, {\em pre-osteoblasts} ({\pb}) and {\em pre-osteoclasts} ({\pc}), as well as by emphasizing affinity and perceptual relations.

Undoubtedly the resulting SP allows a more complex and faithful description at cellular scale than
in \cite{acri2010} - where the rigid spatial approach inherited from CA approximates BMUs as simple spatial grids of $\mbox{O}_c$s without local regulators - but, at the same time, very similar to that one depicted in \cite{cs2bio2010}. For this reason, the proposed SP for bone remodelling can be considered the first step of a formal expressiveness study pivoting on CxA and involving SPs, as well as the shape-based multiscale approach that, previously exploited for bone remodelling in \cite{cs2bio2010}, resulted to be also highly faithful.

\section{The Complex Automata modelling paradigm}
\label{cxa}

The Complex Automata (CxA) \cite{HoekstraFCC08} paradigm has been recently introduced for modelling and simulating multiscale systems and, in particular, the process of development of stenosis in a stented coronary artery \cite{hoedue}.

\begin{figure}
\begin{center}
\includegraphics[height=6.7cm]{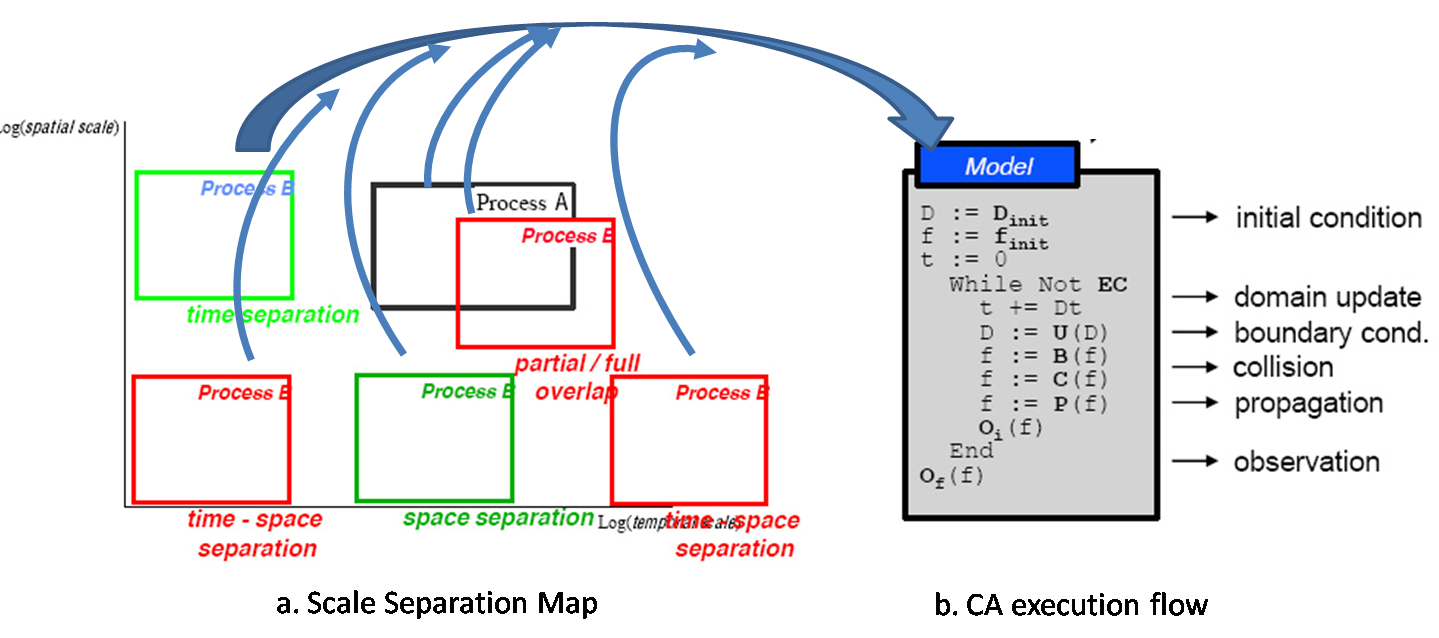}
\end{center}
\caption{a. Scale Separation Map; b. CA execution flow.}
\label{ssmcomp}
\end{figure}

\begin{figure}
\begin{center}
\includegraphics[height=6.9cm]{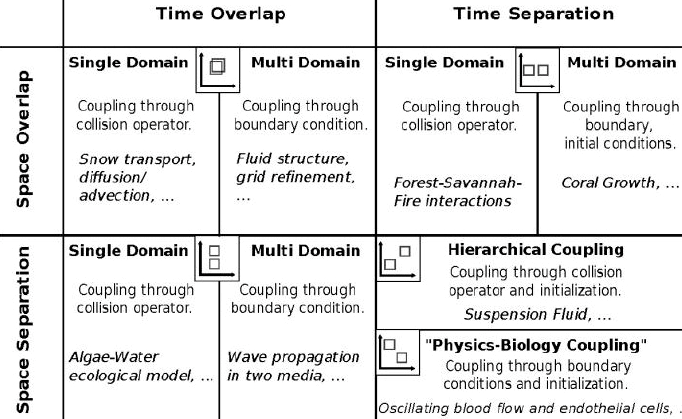}
\end{center}
\caption{SSM and Composition patterns.}
\label{interscalemap}
\end{figure}

In the modelling phase, CxA building blocks are {\em Cellular Automata} (CA) \cite{new66} (i) representing processes operating on different spatio-temporal scales, (ii) characterized by a uniform Lattice Boltzmann Model-like (LBM) update rule - and, as a consequence, execution flow (see Fig.~\ref{ssmcomp} (b)) - (iii) mutually interacting across the scales by well-defined composition patterns\footnote{Due to the lack of space, composition patterns are not discussed here and we refer to \cite{hoepat} for further details.} (see Fig.~\ref{interscalemap}). The simulation phase mainly relies on agent-based models (ABM) \cite{ma1,ma2}, where each ABM corresponds to a specific CA in the CxA model.

More in detail, the update rule of any CA is uniformly defined as a composition of three operators:
{\em boundary condition} $B[\cdot]$ and {\em collision} $C[\cdot]$, both depending on external parameters, and {\em propagation} $P$, depending on the topology of the domain. The $B$ operator is needed to specify the values of the variable that are defined by its external environment (in the case of a LBM fluid simulation, the missing density distributions at the wall). The $C$ operator represents the state update for each cell. The $P$ operator sends the local states of each cell to the neighbors that need it, assuming an underlying topology of interconnection.

In an ABM, the same fundamental operations are also performed with cells replaced by agents. The propagation
procedure sends the local states of each agent to the neighbors that need it.  A special agent can be defined as a centralized information
repository. The data structure is a set of  agents which is traversed in any order because all the above operation are, in nature, parallel operations.

Being the update rule of any CA uniformly defined, such composition patterns only depend on the CA
spatio-temporal ``positions'' in a {\em Scale Separation Map} (SSM), where each CA is represented as an area according to its spatial and temporal scales (see Fig. \ref{ssmcomp} (a)). Formally:

\begin{definition}
\rm
A CxA ${\cal A}$ is a graph $(V,E)$, where $V$ - the set of vertices - and $E$ - the set of edges - are defined as follows:
\begin{itemize}
\item[-] $V= \{C_k \!\stackrel{}{=}^{\!\!\!\!\!\!\mbox{\scriptsize def}}\!\langle (\Delta x_k, \Delta t_k, X_k,  T_k), \St_k, \Phi_k, s^0_k, u_k \rangle | \: C_k \mbox{ is a CA}\}$ where $\forall C_k\in V$,
\par\smallskip - $(\Delta x_k, \Delta t_k, X_k,  T_k)$ denotes the  spatio-temporal domain of $C_k$, i.e.\ $\Delta x_k$ is the cell spatial size, $X_k$ is the space region size, $\Delta t_k$ is the time step and $T_k$ is the end of the simulated time interval of $C_k$;
\par\smallskip - $\St_k$ denote the set of states;
\par\smallskip - $s^0_k\in \St_k$ is the initial state;
\par\smallskip - $u_k$ is a field collecting the external data of $C_k$;
\par\smallskip - $\Phi_k$ is the update rule encoded in LBM style as follows $$s^{n_k+\Delta t_k}_k= P \circ   C[u_k][s^{n_k}_k] \circ B[u_k]$$ where $s^{n_k}_k, s^{n_k+\Delta t_k}_k\in \St_k$ denote resp. the state of $C_k$ obtained as the numerical solution at the $n_k$-th time step and the one  at the $(n_k+1)$-th time step, while $\circ$ denotes, as usual, the operator of function composition.
\item[-] $E= \{E_{hk}| E_{hk} \mbox{ is a {\em composition pattern} between } C_h\mbox{ and } C_k\}$.
\end{itemize}
\end{definition}

Finally, the numerical outcome of each $C_k$ is denoted by $s^{T_k}\in \St_k$.

\subsection{A multiscale trabecular bone remodelling model based on CxA}
\label{cs1}

In the following, we briefly recall the key elements of the model proposed (and fully described) in  \cite{acri2010}. Assuming that $\mbox{O}_y$s act as mechano-sensors, the model - for simplicity proposed in 2D - consists of a CA, whose cells are in turn CAs: the ``macro'' CA (denoted by $C_1$)  models a portion of trabecular bone as a lattice of BMUs (macroscopic slow process), while each ``micro'' CA (denoted by $C_{(i,2)}$, where $i$ corresponds to the cell $i$ in $C_1$) models a single BMU as a lattice of $\mbox{O}_y$s and their surrounding mineralized tissue (microscopic fast process).

The ``macro''  cell size linearly depends on the ``micro'' cell one, which is in turn derived from the $\mbox{O}_y$s estimated density in bone. Assuming that a cubic millimeter of fully mineralized tissue contains 16000 $\mbox{O}_y$s, then a 3D lattice representing this unit volume should contain $25$ ($\approx 16000^{1/3}$) cells in each side. Therefore, a cubic millimeter of bone could be modeled as a 3D lattice of $25^3$ cells, matching with the data reported in \cite{mull96}. As a consequence, a 2D cell lattice with a thickness of $1/25$mm can be structured in $25^2$ cells, matching also with the data presented in \cite{mar90}.

The ``macro'' neighborhood layout can be defined either as the simplest 2D Von Neumann neighborhood ($4$ cells) or as the  2D Moore one ($8$ cells), depending on how  ``local'' we consider the remodelling process on a trabecular region (i.e., in other terms, how ``local'' we consider the propagation of remodelling activation state among BMUs). The ``micro'' neighborhood layout can be defined as the 2D Moore neighborhood.

{\bf Micro execution flow.} The state of each cell $j$ in $C_{(i,2)}$ at a time $t_{(i,2)}$  is defined by its mass fraction $m^j_{(i,2)}(t_{(i,2)})$, varying from $0$ (bone marrow) to $1$ (fully mineralized). The mechanical stimulus $F^j_{(i,2)}(t_{(i,2)})=U^j_{(i,2)}(t_{(i,2)})/m^j_{(i,2)}(t_{(i,2)})$ - being $U^j_{(i,2)}(t_{(i,2)})$ the strain energy density of $j$ at time $t_{(i,2)}$ - is calculated by the {\em Meshless Cell Method} \cite{vic2008} (MCM). Each cell $j$ modifies its mass according to the error signal  $e^j_{(i,2)}(t_{(i,2)})$ between the mechanical stimulus and the internal equilibrium state, determined by the condition $e^j_{(i,2)}(t_{(i,2)}) = 0$; when this condition does not hold, a local collision formula\footnote{The formula can be selected from the approaches presented in \cite{tovar08}} modifies the mass fraction ($m^j_{(i,2)}(t_{(i,2)}+\Delta t_{(i,2)})$) to restore the equilibrium condition. Consequently, the change in mass modifies the stress/strain field in the bone and, therefore, the stimulus operating on $j$. This processes continues until the error signal is zero or no possible mass change can be made. The convergence is satisfied when the change in density is small: if there is no convergence, the process continues with a new MCM analysis.

{\bf Macro execution flow.} Similarly to the micro execution flow, the state of each cell $i$ in $C_1$ is defined by the apparent density $m^i_1(t_1)$, which can vary from $0$ (void) to $1$ (fully mineralized tissue). An homogeneous apparent density distribution for any $i$ corresponds to an
isotropic material, while intermediate values represent trabecular architecture.

A global MCM analysis evaluates the stress field $F^i_1(t_1)$ on $i$ at a time $t_1$, so defining the loading conditions operating on each $i$. We know that $i$ modifies the microstructure by processes of formation/resorption (corresponding to $s^{T_{(i,2)}}$, see below); this process results in formation and adaptation of trabeculae. Hence, the global MCM analysis is  performed over the resulting structure to update the stress field until there is no change in the relative densities and there is no change in the stress field.

\begin{figure}
\begin{center}
\includegraphics[height=5cm]{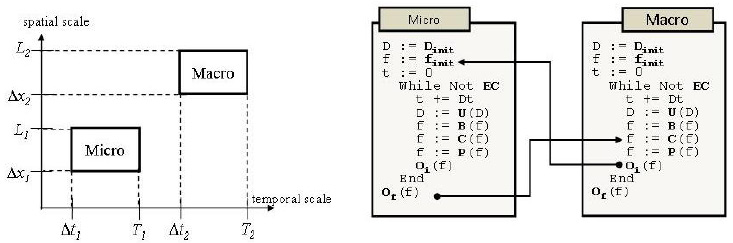}
\end{center}
\vspace{-0.5cm}
\caption{Micro-Macro composition pattern}
\label{mmcase}
\end{figure}

{\bf Micro-Macro composition pattern.} Each $C_{(i,2)}$ is linked to $C_1$ by the ``micro-macro'' composition pattern, defined in Fig.~\ref{mmcase}. More in  detail, $C_1$ takes input from explicit simulations of $C_{(i,2)}$ on each lattice site $i$ at each time step $\Delta t_1$, while each $C_{(i,2)}$ runs to completion, assuming that all $C_{(i,2)}$ are much faster than the macroscopic process and therefore are in quasi-equilibrium on the macroscopic time scale.

A close inspection of this coupling template shows indeed that upon each $C_1$'s iteration each $C_{(i,2)}$ executes a complete simulation, taking input from $C_1$. In turn, each $C_{(i,2)}$ output ($s^{T_{(i,2)}}$) is fed into the $C_1$ collision operator.

\vspace{-0.5cm}

\section{The Spatial P system paradigm}
\label{sp}
Spatial P systems \cite{sps} are a variant of P systems \cite{P02}. Since their introduction, P systems have been widely studied \cite{vienna} as ``biological'' computing devices. A P system contains a hierarchy of membranes - logical compartments that have a defined containment relationship - each of them equipped with a multiset of different objects and a set of evolution rules of the form $u \rightarrow v$. The original biological intuition is that objects represent molecules, membranes represent cell compartments and evolution rules represent biochemical reactions between objects, possibly different in different compartments. At each evolution step a non-deterministic choice is made between possible future states of the system. Each of the possible next states is determined applying the principle of \textit{maximal parallelism}. This means that a state is considered if and only if all possible applicable rules in all membranes are applied, that is to say that no rule exists in a membrane such that the objects $u$ (a string of symbols representing a multiset of objects) that it needs to be activated are still available in the membrane (when a rule is activated then the objects it needs are taken from the available ones in that membrane). When the non-deterministic choice is made, the whole system performs an update according to the applied rules (note that a rule can be applied more than once). The effects are the destruction of the objects taken by the applied (instances of) rules and the generation of the new objects $v$ created by the rules sent (1) in the same membrane, (2) in one of the immediately inner membranes or (3) out of the surrounding membrane. The skin membrane is the one that surrounds all the hierarchy.

The different structure of rules (cooperating or not), the number of membranes or the number of objects used, the possibility of defining priority between rules and the possibility to dissolve some membranes are only some of the studied features of P systems w.r.t. their computability power (they are indeed Turing equivalent) and their capabilities to represent different biological scenarios. For a more formal and comprehensive description of this formalism we refer to \cite{P02,vienna} and references therein.

Let us now introduce Spatial P systems (SP) at the level of detail needed for the purpose of this paper. For a more formal introduction we refer to \cite{sps}. SPs extend P systems by embedding membranes and objects into the two-dimensional space with natural coordinates $\bbbn^2$. Membranes have rectangular shape and, as for normal P systems, can be nested. The spatial description of a membrane is given in terms of (i) the position $p \in \bbbn^2$ of its bottom-left corner w.r.t. the parent membrane, (ii) the membrane extents along the two dimensions, i.e.\ its width $w$ and height $h$, $w,h \in \bbbn^+$. There is always a distinguished skin membrane, which contains all other membranes and objects. The skin membrane is assumed to be labeled with $1$ positioned in $(0,0)$ with respect to the global coordinate system.

\begin{figure}
  \begin{center}
    \includegraphics{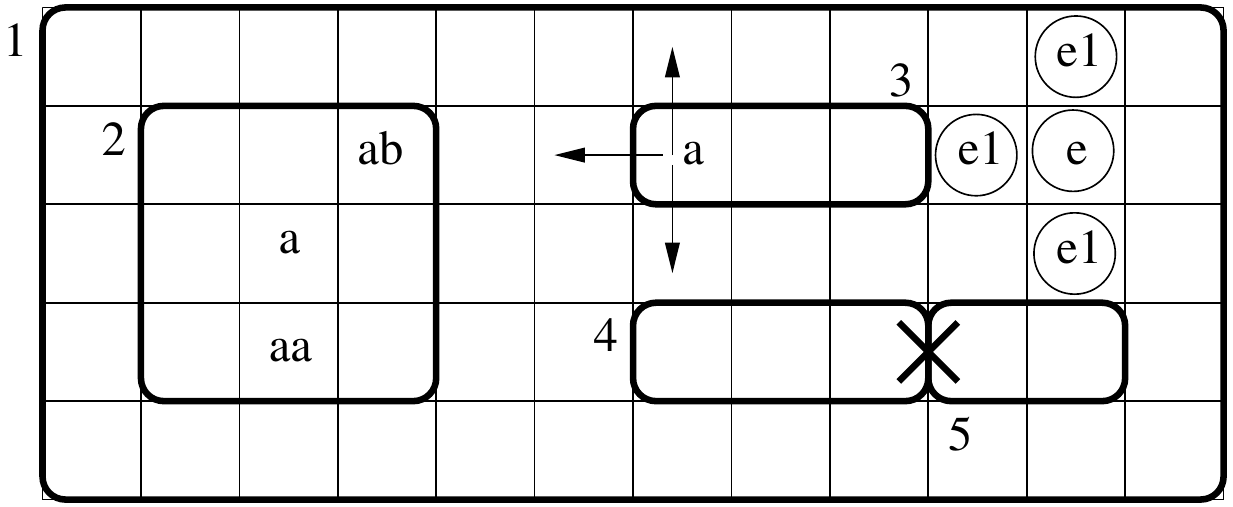}
  \end{center}
\caption{An example of a Spatial P system with some objects, some mutually exclusive objects and two wrongly located membranes.}
\label{fig:sp1}
\end{figure}

An example of Spatial P system is shown in Fig.~\ref{fig:sp1}. Membrane $1$, having width $12$ and height $5$, contains four membranes labeled $2,3,4$ and $5$. The spatial extension of membrane $2$ is described by its bottom-left corner in $(1,1)$, its width $3$ and height $3$. For membrane $3$, its position is $(6,3)$, and its dimensions are $(3,1)$. Four objects $a$ are contained in membrane $2$, two at positions $(1,0)$, one at $(1,1)$ and one at $(2,2)$, where also an object $b$ resides. Membrane $3$ contains an object $a$ in position $(0,0)$. Membrane $1$ contains four mutually exclusive objects: three objects $e_1$ in positions $(10,4),(9,3)$ and $(10,2)$ and one object $e$ in position $(10,3)$. All other positions, in all regions, are empty.

Nesting of membranes has to satisfy some intuitive spatial constraints: sibling membranes must not overlap and membranes cannot exceed the bounds of their parent membranes. Moreover, we do not allow membrane edges to be \emph{adjacent} (for instance, membranes 4 and 5 of Fig.~\ref{fig:sp1} are adjacent and not allowed).

The membrane structure is a partition of the space bounded by the skin membrane. A position belongs to a membrane if and only if it is contained within its bounds and not contained in any other child membrane. The set of all positions belonging to a membrane is called a \emph{region}. For example, the region of membrane $2$ of Fig.~\ref{fig:sp1} is the set $reg(2) = \{(x,y) \mid 1 \leq x,y \leq 3\}$.

Each object in a Spatial P system model is located at a position in the region of a membrane. There are two kinds of objects, \emph{ordinary} objects and \emph{mutually exclusive} (\textit{ME}) objects, which are represented by two disjoint sets $V$ and $E$, respectively. The difference between them is that two {\me} objects are not allowed to occupy the same position at the same time, while any number of ordinary objects can be positioned at the same cell. In Fig.~\ref{fig:sp1}, there are three {\me} objects $e_1$ and one {\me} object $e$. They are represented with a little circle around them in the cell. In this case, the rule imposes that the object $e$ could not stay at any of the positions of the objects $e_1$.

A set of evolution rules is associated with each membrane. Evolution rules are either of the form (i) $u \to v$ or (ii) $u_1 - u_2 \to v_1 - v_2$, where $u, u_1$ and $u_2$ are strings over objects, and $v, v_1$ and $v_2$ are strings of messages (where each message is composed of a multiset of objects, and a target indication). A rule of the form $u \to v$ is meant to be applied to each position forming the membrane region, provided that all the objects $u$ appear in that position. String $v$ specifies the products of the rule and their resulting positions.
A rule of the form $u_1 - u_2 \to v_1 - v_2$ describes a simultaneous application of two rules, $u_1 \to v_1$ and $u_2 \to v_2$, to two \textit{adjacent} positions inside the membrane region. Two positions are adjacent if and only if their Manhattan distance is exactly $1$\footnote{The Manhattan distance between two positions $(x_1,y_1),(x_2,y_2) \in \bbbz^2$ is $\lvert x_1 - x_2\rvert + \lvert y_1 - y_2\rvert$.}.

Rules can send an object either into an inner membrane or out of the membrane. Moreover they can  specify a displacement for the objects that remain in the same membrane (thus any velocity is allowed within the same membrane). Messages are of the following forms:
\begin{itemize}
\item[-] $v_{\delta p}$, with $\delta p \in \bbbz^2$, the multisets of objects $v$ are added to position $p + \delta p$ relative to the position $p$ in which the rule is applied;
\item[-] $v_{out}$, the multisets of objects $v$ are to be sent out of the membrane;
\item[-] $v_{in_l}$, the multisets of objects $v$ are to be sent into the child membrane identified by $l$.
\end{itemize}
Note that there is not an explicit target $here$, used in conventional P systems for messages of the form $v_{here}$. In Spatial P systems target $here$ means ``in the current position'' (inside membrane region), and thus it has a narrower meaning than in conventional P systems. It can be defined as an alias for the null position $here = (0,0)$. We also use the following abbreviations for the relative positions $\delta p$ denoting adjacent positions: $N = (0,1)$, $S = (0,-1)$, $E = (1,0)$, $W = (-1,0)$\footnote{North, South, East, West.}.

For example, the rule $A \to (b)_{(2,0)}\, (c)_{out}\, (d)_{in_{2}}$ can be applied to an object $A$, which results in an object $b$ in position $p + (2,0)$ relative to the current position $p$; an object $c$ being sent out of the membrane; and an object $d$ being sent into the inner membrane labeled $2$.

Let $TAR$ denote the set of message targets $\bbbz^2  \cup  \{out\} \cup  \{in_i \mid i \in \bbbn\}$. Given a set of objects $O$ we denote with $O_{tar}$ the corresponding set of messages $O \times TAR$. Hence, in evolution rules (i) $u \to v$ and (ii) $u_1 - u_2 \to v_1 - v_2$, we have $u,u_1,u_2 \in (V \cup E)^{*}$ and $v,v_1,v_2 \in ((V \cup E)_{tar})^{*}$. A \emph{Spatial P system} is a tuple $ (V, E, \mu, \sigma, W^{(1)} , \ldots , W^{(n)}, R_1, \ldots, R_n)$ where $V$ and $E$ are disjoint \emph{alphabets}, $\mu$ is a description of the tree-structure of membranes, $\sigma$ is a description of the spatial distribution of membranes, $W^{(i)} = \{ w^{(i)}_{x,y} \}$ with $1 \leq i \leq n$ are sets of strings $w^{(i)}_{x,y} \in (V \cup E)^{*}$ each string $w^{(i)}_{x,y}$ representing a multiset over $V \cup E$ associated with position $(x,y)$ inside membrane $i$, $R_i$ is the finite sets of \emph{evolution rules} associated with membrane $i$.

Given a position $p = (x,y)$ in membrane $i$, an evolution rule $u \to v$ is \emph{$p$-enabled} iff
(i) whenever the rule specifies any $out$ target, then $p$ is adjacent to an edge of membrane $i$; (ii) whenever the rule specifies a target $in_j$, then $j$ is a child membrane of $i$, and $p$ is adjacent to it; (iii) for any target position $\delta p$, the resulting position $p' = p + \delta p$ with respect to the current position $p$ is contained in membrane region. A multiset of evolution rules of the form $u \to v$ is \emph{applicable} to a position $p$ inside a region of the system iff each evolution rule is $p$-enabled, and all reactant objects (with their multiplicities) are present in $p$.

In each step of the evolution of a Spatial P system, some evolution rules are chosen according to the principle of maximal parallelism, just as in the case of P systems. However, some additional restrictions, regarding spatial consistency, apply. First, the chosen multisets of rules must be applicable on the whole, namely for each position $x$ in membrane region, the multiset of rules $M^{(x)}_1 \cup \{u_1 \to v_1 \mid \exists q.\; u_1-u_2 \to v_1-v_2 \in M^{(x,q)}_2\} \cup \{u_2 \to v_2 \mid \exists q.\; u_1-u_2 \to v_1-v_2 \in M^{(q,x)}_2\}$ must be applicable. Moreover, they are required to be \emph{valid}, namely two {\me} objects are forbidden to end up occupying the same position $p'$ at the end of the step\footnote{However, note that, during the step, a {\me} object can disappear from a position and another one can take its place.}.

There are also some additional sources of non-determinism. For a message $v_{in_j}$, the objects are placed in the nearest position $p'$ in the region of $j$, with respect to the current position. For a message $v_{out}$, the objects are placed in one of the nearest positions outside the membrane. In case the output position for a $v_{out}$ message is not unique, as is the case when the rule is applied to a vertex position, then the output position is chosen non-deterministically along the horizontal and vertical direction. In Fig.~\ref{fig:sp1} the object $a$ in membrane $3$ can end up in one of the three positions indicated by the arrows if it has been sent out of the membrane by a rule.

The definitions of \emph{computation} and {successful} computation from standard P systems also apply to Spatial P systems (see \cite{sps}). However, for our purposes, an approach to termination more close to the simulation world, w.r.t. a termination condition suitable for showing the universality of the formalism, can be taken. Thus we say, for simplicity, that a given P system terminates whenever it performed a predetermined number of evolution steps $\mathtt{MAX\_SIM}$. This simplification can, in any case, be removed to obtain a more reliable model in which every Spatial P system has to reach a sort of ``equilibrium'' condition (similar to the termination defined in \cite{sps}) to terminate.

\subsection{A multiscale trabecular bone remodelling model based on SP}
\label{cs2}
Following the approach of CxA presented in Section~\ref{cxa} we define a micro-macro coupling scheme between two different scales (tissue and cellular/BMU) of the bone remodelling phenomenon for which a CxA model has been presented in Section~\ref{cs1}. As discussed in the introduction, in this case the lower level model (BMU) is richer because it fully exploits the native spatial aspects and individual-based nature of Spatial P systems.

{\bf Macro model}. To implement with SPs the micro-macro composition pattern we define a SP $S_1$ representing the tissue level. It consists of only the skin membrane. Making all the simplifications described in Section~\ref{cs1}, we consider a square grid of about $25^2$ cells. Each of them contains, at any moment, a number of objects $c$ proportional to their mineralisation value (expressed as a density in a certain interval). Moreover, depending on the mineralisation of every cell, an ``activator object'' $a$, inserted in every cell, starts a process that determines if the cell is on the surface of the bone. The evolution rules are the following: $$c^ma \rightarrow b_1d_1 \quad c^nb_1 \rightarrow c^{n+m}b \quad d_1 \rightarrow d \quad db \rightarrow \lambda\footnote{$\lambda$ is the empty string.} \quad db_1 \rightarrow c^mf$$ A cell is on the surface iff the number of `$c$'s is in the interval $[m, m+n)$. This interval models the given threshold above which a cell is considered fully mineralized (here one object $c$ represents a certain amount of density that is taken small enough to model the threshold with the wanted precision). The given rules are such that iff after 2 evolution steps an object $f$ is present in a cell, than that cell is on the surface of the bone (this means that it is possibly subject to remodelling). After this check, some of the cells will be surely selected if an object $g$, representing a microdamage, is present in the cell (objects $g$ are distributed initially over the grid according to the forces applied to the tissue and the current mineralisation). If not, a cell with an $f$ could still be activated if selected randomly for that. For this purpose an object $h$ is inserted at the beginning on a certain percentage of the cells. Concluding, the third step of evolution of $S_1$ will apply one of these other rules: $$fg \rightarrow r \quad fh \rightarrow r$$ The presence of an object $r$ in a cell after the third evolution step signals the activation of remodelling on that cell. Note that in rule-based systems like SPs the modelling of a simple concept like sequentiality or conditional choice can be tricky, as the above example shows. Indeed, each paradigm has its own strong points, but also its weakenesses.

\begin{figure}
  \begin{center}
    \includegraphics{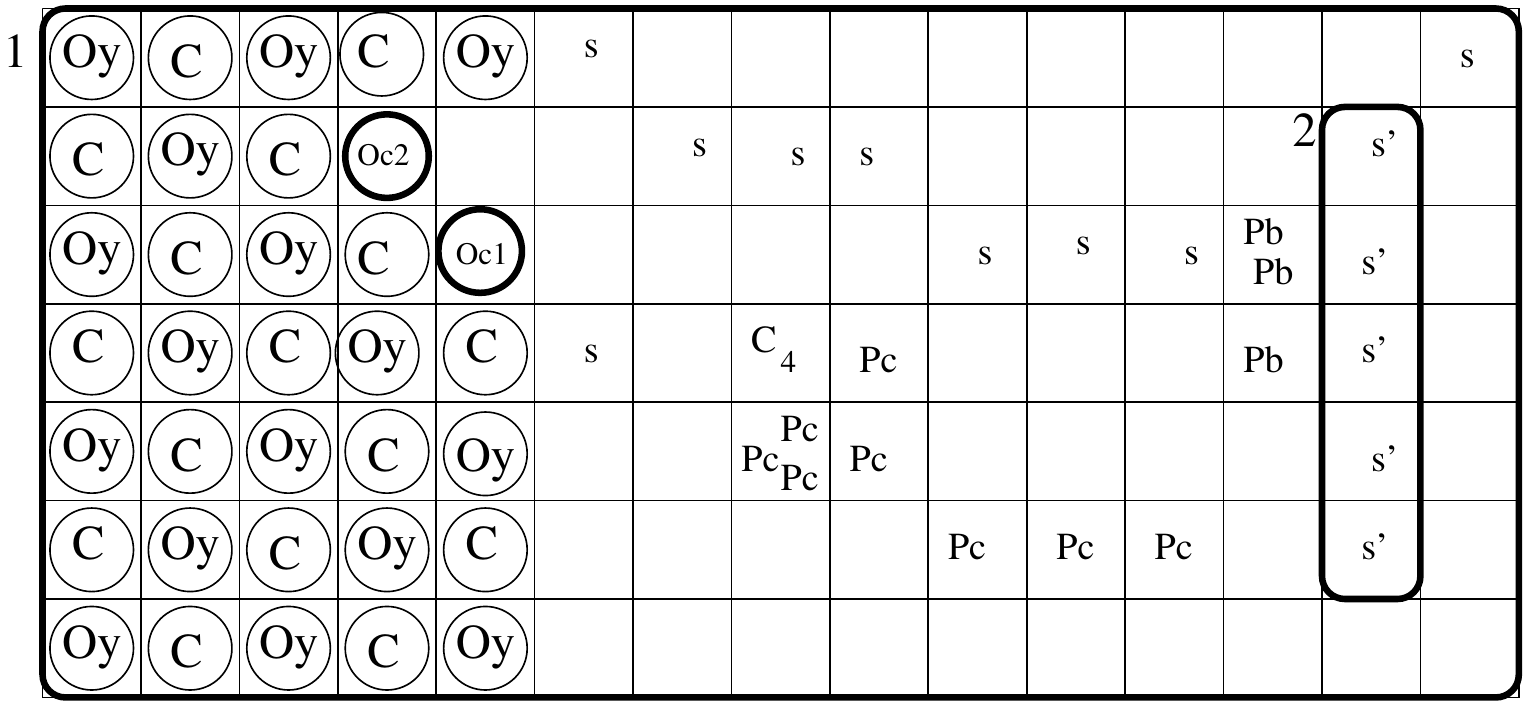}
  \end{center}
\caption{A Spatial P system representing a BMU activated for bone remodelling. In the current phase some osteoclasts have already formed and started to destroy the upper part of the surface of the mineralized part.}
\label{fig:spbmu}
\end{figure}

{\bf Micro model}. For every cell $i$ of $S_1$ we define a corresponding SP $S_{(i,2)}$ representing a BMU associated to that portion of space. In this case the structure is more complex. Fig.~\ref{fig:spbmu} shows a scenario in which a system $S_{(i,2)}$ has been activated for remodelling and is in an intermediate state of its evolution. If a BMU is activated for remodelling it contains, at the beginning, a particular starter object $s$ in a certain cell. Note that if such an object is not present no evolution rule can be applied making $S_{(i,2)}$ to run silently for the given $\mathtt{MAX\_SIM\_BMU}$ number of evolution steps that are to be specified for the lower level process to be completed. If the BMU is activated, the initial configuration depends on the actual degree of mineralisation (given by the higher level). In any case, the grid is divided into two zones: the left one is mineralized and the right one is not. The mineralisation is represented by {\me} objects \oy\ (representing a little portion of bone containing an osteocyte) or $C$ (the same portion, but not containing an osteocyte).

Note that on the right side of the grid there is an inner membrane $2$. This membrane represents the link of the BMU with blood and the bone marrow. From it all the cells necessary to the process will come out at the proper times, ``called'' by the objects already present in the non-mineralized part. The signal $s$ has to spread over the non-mineralized part moving towards East until it finally reaches membrane $2$ in which it enters. To simulate this we have the rules (in membrane 1): $$s \rightarrow s_N s_E s_S \quad s \rightarrow s_E \quad s \rightarrow s_{{in}_2} \quad s \rightarrow s_{out}$$

\noindent
As soon as the first $s$ enters membrane $2$\footnote{They all eventually end up into membrane $2$ or are expelled out of the skin membrane.} a certain number of pre-osteoclasts \pc\ and pre-osteoblasts \pb\ are released in the non-mineralized zone of membrane $1$ from membrane $2$. The rules to make this are the following (in membrane $2$): $$s \rightarrow s'(\mbox{\pc})^k_{out}(\mbox{\pb})^\ell_{out} \quad s's \rightarrow s' \quad s' \rightarrow s's'_N \quad s' \rightarrow s's'_S$$ Here $k$ and $\ell$ are parameters of the model (like the threshold, expressed with $n$ and $m$, at the tissue level). Note that all the signals $s$, after the first one, that eventually end up in membrane $2$ are destroyed by $s'$, which replicates and spreads itself all along membrane $2$.

The pre-osteoclasts \pc\ and the pre-osteoblasts \pb\ are now free in the non-mineralized space. The \pb s, at this stage, do not move, while \pc s have to start a process of aggregation in order to form a full \oc. To realize this we make \pc s move randomly around with the possibility to start an aggregation as soon as 4 of them are close to each other. An object $C_n$, with $4 \leq n < N\_OC$, continuously absorbs \pc s in its neighbourhood (counting one for each of them) until it reaches the number $N\_OC$, a parameter of the model indicating at which number of aggregated \pc s the aggregation becomes a fully grown osteoclast. The rules are the following (in membrane $1$): $$\mbox{\pc} \rightarrow \mbox{\pc} \quad \mbox{\pc} \rightarrow \mbox{\pc}_N \quad \mbox{\pc} \rightarrow \mbox{\pc}_S \quad \mbox{\pc} \rightarrow \mbox{\pc}_O \quad \mbox{\pc} \rightarrow \mbox{\pc}_E \quad \mbox{\pc}^h \rightarrow C_h \quad \mbox{\pc}^{h_1}-\mbox{\pc}^{h_2} \rightarrow \lambda-C_{h_1 + h_2}$$ $$C_h-\mbox{\pc} \rightarrow C_{h+1}-\lambda \quad C_h \mbox{\pc} \rightarrow C_{h+1} \quad C_{N\_OC - 1}-\mbox{\pc} \rightarrow \mbox{\oc}_0-\lambda \quad C_{N\_OC - 1} \mbox{\pc} \rightarrow \mbox{\oc}_0$$ where instances of the rules with variables $h, h_1$ and $h_2$ are present for all values such that $4 \leq h < N\_OC$, $4 \leq h_1 + h_2 < N\_OC$.

After an \oc\ is formed it starts moving towards the mineralized zone to destroy the existing bone. To simulate this, we make them move towards West\footnote{Another modelling choice could be that \oc s move around randomly. In this case, however, we chose to model a gradient of attraction towards the mineralized part by forcing them to move only to the West direction.} until they cannot proceed because another {\me} object (a mineralized cell with or without osteocyte) is present. In this case they destroy it and count one. They continue to destroy towards West until they destroyed a given number of cells $N\_DC$ at which they decide to die. Their death starts another signal $o$ that, similarly to the first $s$, will trigger the formation of osteoblasts for bone reconstruction. The rules are:  $$\mbox{\oc} \rightarrow \mbox{\oc}_W \quad  \mbox{\oy}-\mbox{\oc}_z \rightarrow \mbox{\oc}_{z+1}-\lambda$$ $$C-\mbox{\oc}_z \rightarrow \mbox{\oc}_{z+1}-\lambda \quad \mbox{\oy}-\mbox{\oc}_{N\_DC -1} \rightarrow \lambda-o \quad C-\mbox{\oc}_{N\_DC -1} \rightarrow \lambda-o$$ Note that, in this case, a delay can be introduced in order to represent the fact that an \oc\ takes some time to destroy a portion of the bone.

For the sake of brevity, we do not continue the description of the other phases of the bone remodelling. The reader can easily figure out how they proceed. Moreover, it is easy to see that this model can be enriched with more details and made more complex using the qualitative and quantitative information available in the literature about the bone remodelling phenomenon. Our purpose is just to show that such a model can be constructed and coupled with the higher level model. In the following we explain the coupling in more details.

{\bf Integration scheme}. We use coupling functions $f\downarrow$ and $f\uparrow$ to integrate the two models at the two considered scales. Function $f\downarrow$ says, for each cell $i$ of $S_1$, the initial configuration of the Spatial P system $S_{(i,2)}$. In particular, if the object $r$ is present at cell $i$ in $S_1$ it will make the object $s$ available in some cell of $S_{(i,2)}$. Moreover, considering the mineralisation of cell $i$ represented by the number of $c$ objects in there, it determines how many cells on the left side of $S_{(i,2)}$ contains {\me} objects representing the mineralized zone. Conversely, $f\uparrow$ tells, considering the whole $S_{(i,2)}$ state after $\mathtt{MAX\_SIM\_BMU}$ evolution steps have been performed, how many $c$ objects must be placed on cell $i$ of $S_1$. The whole coupled process is as follows:
\begin{enumerate}
  \item Initialize $S_1$ putting in each cell the initial number of $c$ objects and the activator $a$
  \item \label{goto} Determine, using the model of forces and the data on mineralisation, in which cells of $S_1$ objects $g$ and $h$ have to be placed
  \item Perform three evolution steps of $S_1$
  \item Apply $f\downarrow$ from each cell $i$ of $S_1$ to each $S_{(i,2)}$
  \item \label{downsteps} Perform $\mathtt{MAX\_SIM\_BMU}$ evolution steps on all $S_{(i,2)}$
  \item Apply $f\uparrow$ from each $S_{(i,2)}$ to each cell $i$ of $S_1$
  \item Goto step~\ref{goto}.
\end{enumerate}
To conclude, we want to underline that the simplification about the maximum number $\mathtt{MAX\_SIM\_BMU}$ of evolution steps can be simply overcome by defining a termination condition for the lower level Spatial P systems in order for them to reach a final state of the process in the BMU. In this case, however, two different $S_{(i,2)}$ and $S_{(j,2)}$ could take different times to complete step~\ref{downsteps}.\ of the above process. This could make the coupling a little bit more complicated to implement.

\section{Conclusion and future work}
\label{fw}
The SP paradigm has been here exploited to define a uniform multiscale model for bone remodelling, taking inspiration from the CxA proposed in \cite{acri2010}. Scale-independence property and ability of expressing spatial information are altogether elements which heavily draw up both modelling approaches. In this perspective, we  plan to stress SPs in the attempt of implementing  other CxA composition patterns, i.e.\ taking into account other multiscale scenarios where scales are not necessarily related according to a micro-macro scheme.

SPs, differently from CxA, are able to handle an infinite number of different ``states'' because in  each cell any number of different objects can be locate. Moreover, localized update rules seem to be more expressive than the single deterministic update rule of CxA. As a consequence, a formal study of the expressive power of the above modelling approaches is under investigation.

Finally, since the highly detailed cellular view of bone remodelling here proposed in term of SP is very close to that one depicted in \cite{cs2bio2010} in terms of shapes, we believe that such an expressiveness study must also include, maybe pivoting on CxA, the shape-based approach described in \cite{iccs2010,cs2bio2010}.

\bibliographystyle{eptcs} 
\bibliography{MultiscaleSPbiblioTris}
\end{document}